\documentstyle[aps,prl,epsf]{revtex}
\begin{document}
\draft

\twocolumn[\hsize\textwidth\columnwidth\hsize\csname@twocolumnfalse%
\endcsname

\title{Lateral Separation of Macromolecules and Polyelectrolytes 
in Microlithographic Arrays}
\author{Deniz Erta\c s} 
\address{Lyman Laboratory of Physics, Harvard University, Cambridge, 
Massachusetts 02138}

\date{\today} 

\maketitle

\begin{abstract}
A new approach to separation of a variety of microscopic and mesoscopic
objects in dilute solution is presented. The approach takes advantage 
of unique properties of a specially designed separation device
(sieve), which can be readily built using already developed
microlithographic techniques\cite{Austin}.
Due to the broken reflection symmetry in its design, the {\it direction} 
of motion of an object in the sieve varies as a function of its 
self-diffusion constant, causing separation 
transverse to its direction of motion. This gives the device some 
significant and unique advantages over existing fractionation methods 
based on centrifugation and electrophoresis.  
\end{abstract}
]

Separation of macromolecules such as proteins and DNA, as well as mesoscopic
objects such as cells and latex spheres according to size has 
many important technological applications, and an immense effort has gone 
into achieving efficient, well controlled and high resolution separation 
techniques\cite{Austin,Gaal,Birnie}. Two of the main avenues that have 
been employed extensively are electrophoresis\cite{Gaal} and 
centrifugation\cite{Birnie}. 
These two methods are somewhat complementary since the former typically 
separates by charge density and the latter by mass density. 
In particular, virtually all electrophoretic techniques 
rely on an electrophoretic mobility $\mu_e$ that changes as a 
function of molecular weight $M_{\rm w}$, or some other characteristic
for which separation is desired, since the separation occurs along 
the same direction as the average motion. An initially polydisperse band 
separates into many bands containing objects of different sizes as they travel 
at different velocities ${\bf v}(M_{\rm w})=\mu_e(M_{\rm w}){\bf E}$
along the direction of the applied field $\bf E$.
A major obstacle to the electrophoretic 
separation of large polyelectrolytes with constant charge density such as 
nucleic acids is the independence of their electrophoretic mobility to their
molecular weight in solution. To achieve separation, the polyelectrolytes are 
typically placed in a gel medium, where steric interactions generate a
size-dependent $\mu_e$. Despite significant progress in refined gel
electrophoresis techniques, large objects such as cells or subcellular
structures are impossible to separate due to the limited range of achievable
pore sizes, and issues such as sample loading and recovery are especially 
problematic for fragile specimens\cite{Gaal}.

In this paper, a completely new approach to separation, which embodies 
the advantages of both free flow electrophoresis\cite{Hannig} and gel 
electrophoresis, and is made possible by microlithographic techniques 
recently introduced by Volkmuth and Austin\cite{Austin}, is presented. 
The general idea is to design
a particular electrophoresis chamber (sieve) such that the electrophoretic 
mobility tensor is non-diagonal, i. e., the objects
do not move along the direction of applied electric field {\bf E}. 
Furthermore, the direction of motion varies as a function of size, causing
objects of different sizes to move along different directions.
As a result, this sieve causes lateral separation as in free flow 
electrophoresis\cite{Hannig} without the complications involved in
creating a uniform transverse laminar flow field. The sieve is reusable, 
and the approach enables continuous operation since 
the paths are separated spatially rather than temporally, making retrieval
of separated products extremely easy. These properties potentially 
provide a significant advantage over traditional methods for separation 
of large quantities.

In the remainder of this Letter, a specific realization of such a sieve is 
presented. The dynamics of a polyelectrolyte, e. g., a DNA molecule, 
subject to a 
uniform electric field in the sieve, and the mobility and resulting separation
characteristics are derived. These results are then verified by 
numerical simulation, followed by a detailed comparison of the method
with existing separation techniques.

\begin{figure}
\narrowtext
\centerline{\epsfxsize=3.1truein\epsffile{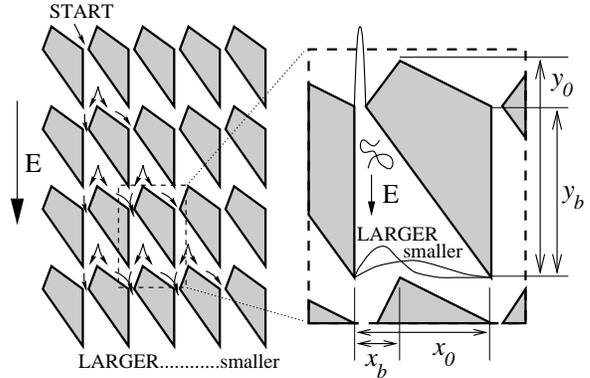}}
\medskip
\caption{The geometry of the sieve. Molecules enter the top of a cell
from a narrow opening and diffuse away from the left wall as they are pulled
down by the electric field. Smaller molecules diffuse farther and are therefore
more likely to end up to the right of the branching point, located at a 
distance $y_b$ from the entrance and $x_b$ from the left wall.
Therefore, the branching probability $p_{\rm B}$, and subsequently 
the macroscopic mobility of a molecule, depends on its size.}
\label{geometry}
\end{figure}

The geometry of the sieve is shown in Fig.~\ref{geometry}. It consists of
a rectangular array of ``cells'' of periodicity $x_0 \times y_0$ 
that have a narrow entrance at the top and two exits at the bottom,
which connect to the next row of cells.
DNA fragments, characterized by a persistence length $\ell_p$, contour 
length $L=N\ell_p$, and diameter $d$, enter from the top and move down 
the cell subject to an electric field ${\bf E}=E{\bf e}_y$. 
The radius of gyration $R_g$ and the self-diffusion constant $D_G$ of
a DNA fragment of $N$ persistence lengths are given by
\begin{eqnarray}
R_g&\simeq&\ell_p N^\nu, \\
D_G&\simeq&D_0 N^{-\alpha},
\end{eqnarray}
where $\nu=1/2$ is the swelling exponent, $D_0$ is the diffusion
constant for a single persistence length,
and $\alpha=1$ for the Rouse model\cite{Rouse} (free draining conditions) 
and $\alpha=\nu=1/2$ for the Zimm model\cite{Zimm}, where
hydrodynamic interactions are taken into account\cite{Doi}. 
The electrophoretic mobility of the fragments is independent of size and 
given by\cite{Ajdari}
\begin{equation}
\mu_e=\frac{\lambda}{\zeta},
\end{equation}
where $\lambda$ and $\zeta$ are the effective charge and friction 
coefficient of a persistence length of the 
DNA, respectively. (The Reynolds number is 
very small in all cases of interest, and inertial effects can be ignored.)
Note that hydrodynamic interactions induced by Brownian Motion are not 
screened by counter-ions, as is the case of electrophoretic velocity 
fields\cite{Russel,Barrat}.
Consider a fragment that enters a cell at time $t_0$ and diffuses away from 
the left wall as it drifts down the cell.
Ignoring its internal modes and characterizing its dynamics
simply by its electrophoretic mobility $\mu_e$ and self-diffusion
constant $D_G$, the probability of finding the DNA a distance $x$
away from the left wall at time $t$ when it has drifted a distance 
$y(t)=\mu_e E (t-t_0)$ from the top of the cell is given by
\begin{equation}
\label{eqpcom}
P_{\rm COM}(x : y)\simeq\frac{\mu_e E x}{2 D_G y}
\exp\left(-\frac{\mu_e E x^2}{4 D_G y}\right),
\end{equation}  
which can be obtained from the solution to the diffusion equation with
reflective (Dirichlet) boundary conditions on the left wall, and ignoring
the effect of the right wall. 
A branching point is located at a distance $x_b$ from the left wall 
and $y_b$ from the top of the cell. Fragments that have
diffused farther from the left wall than the branching point
are funneled to the entrance of 
the cell diagonal to the original one, whereas those that are
closer to the wall end up at the entrance to the cell immediately below.
The probability of branching can be calculated from Eq.(\ref{eqpcom})
as
\begin{eqnarray}
p_{\rm B}&=&\int_{x_b}^{\infty} dx\,P_{\rm COM}(x : y_b)  \nonumber \\
\label{eqhistogram}
&\simeq&\exp\left(-\frac{\mu_e E x_b^2}{4 D_G y_b}\right).
\end{eqnarray}
For a DNA fragment of $N$ persistence lengths, 
\begin{equation}
\label{eqpb}
p_{\rm B}(N)\simeq A e^{-(N/N_0)^\alpha}, 
\end{equation}
where
\begin{equation}
N_0\simeq\left(\frac{4D_0 y_b}{\mu_e E x_b^2}\right)^{1/\alpha}
\end{equation}
is the characteristic separation size of the sieve, and $A$ is a constant 
fitting parameter of $O(1)$. A remarkable observation is that $N_0$ can be 
tuned simply by changing the applied electric field, increasing the dynamic 
range of separation dramatically.

A more accurate analytical estimate of $p_{\rm B}(N)$ 
requires taking into account internal relaxation of the polymer, effective 
wall potentials at distances less than the radius of gyration of the polymer, 
corrections due to diffusion along the field direction, among other effects.
For example, the right wall can no longer be ignored for
$N < \left(\frac{\mu_e E x_0^2}{4D_0 y_b}\right)^{1/\alpha}$,
and the finite size of the fragment becomes significant when
$R_g>x_b$, or equivalently $N>(x_b/\ell_p)^{1/\nu}$. Relaxation effects will
be important when cell traversal time $y_b/(\mu_e E)$ is less than
the principal relaxation time $R_g^2/D_G$, i. e., for 
$N>\left(\frac{\mu_e E \ell_p^2}{D_0 y_b}\right)^{1/(\alpha+2\nu)}$.
Computation of these effects is beyond the scope of this Letter.
However, from a practical standpoint, characterization for a given
geometry can be more readily achieved by numerical simulation
or experiment for most applications, since $p_{\rm B}(N)$ completely 
characterizes the separational properties of the sieve and the periodic 
structure greatly simplifies the numerical determination of this quantity
by simulating a single cell.

At this point, it is useful to point out that any sieve design that breaks
the left-right symmetry can in principle be used for purposes of
separation. The underlying idea is reminiscent of 
rachet potentials\cite{Curie,Magnasco}, which have also been recently 
proposed as particle separators\cite{Prost,Slater}. 
Both rachets and the sieve exploit differences in diffusion constants
and steric constraints, and if the $y$ axis is identified with time, 
the sieve can actually be thought of as the time history of a rachet 
potential. However, unlike rachets, the sieve does not require 
time dependent potentials and is able to operate in continuous mode.

\begin{figure}
\narrowtext
\centerline{\epsfxsize=2.9truein\epsffile{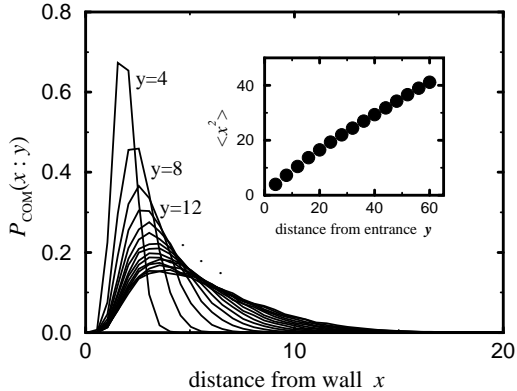}}
\medskip
\caption{Evolution of the density profile of chains with $N=24$ persistence
lengths during the traversal of a cell. After an initial relaxation period,
the profile approximately takes the form given by 
Eq.~(\protect\ref{eqhistogram}), and the mean square distance from the 
left wall increases linearly as a function of the distance from the entrance,
as demonstrated in the inset.
In this run, the primitive cell size of the rectangular lattice is 
$x_0=40 \times y_0=80$; $y_b=60$ and $x_b=4.8$ (all in arbitrary units.).}
\label{evolution}
\end{figure}

\begin{figure}
\narrowtext
\centerline{\epsfxsize=2.9truein\epsffile{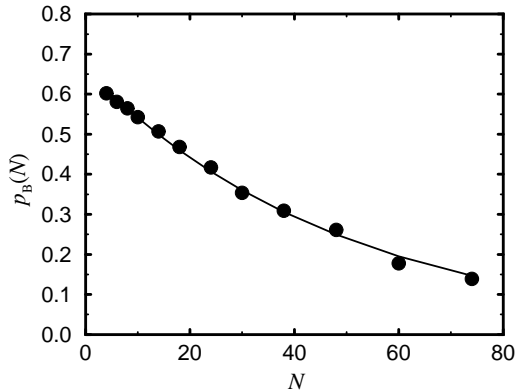}}
\medskip
\caption{Branching probability as a function of the number of segments 
$N$. The line is a fit to the exponential form (\protect\ref{eqpb}) 
with $\nu=1$ and $N_0\approx49$.}
\label{pb}
\end{figure}

Numerical integration of the equations of motion using the Rouse Model
for polymers\cite{Rouse,Doi} $(\alpha=1)$ and reflecting boundary 
conditions at the walls is in good agreement with analytical results. 
Figure~\ref{evolution} shows how the probability density for the 
COM position of a polymer with $N=24$ evolves as it moves down the cell. 
After the initial internal relaxation time, the probability can be fitted 
to the form of Eq.(\ref{eqhistogram}); the inset shows the expected linear 
increase in the mean square distance $\langle x^2 \rangle$ from the left 
wall as a function of distance $y$ from the top of the cell. 
Figure~\ref{pb} shows the branching probability $p_{\rm B}(N)$ as a function
of polymer size, and has the expected exponential behavior, even though
the sizes of the polymers become quite large compared to the cell feature
size. A correlation analysis of the time series of branching events
confirms that they are statistically independent from one 
cell to the next. 

Although this method of separation sounds very promising in principle, it is
important to assess performance parameters and feasibility before a
decision can be made about its practicality and whether it can compete
against established techniques for certain tasks. One of the most important
issues is the resolution that can be achieved\cite{Lerman}. 
A monodisperse packet of 
polymers with size $N$ will spread laterally as it moves through the sieve, 
and after passing through $M$ rows, the density profile of te band will
exhibit a Bernoulli distribution whose peak is located at
$X_P(M)=x_0 M p_{\rm B}(N)$ and whose variance is
$\sigma^2(N)=x_0^2 M p_{\rm B}(N)[1-p_{\rm B}(N)$\cite{Chandra}.
Hence, the full width at half maximum (FWHM) of the corresponding band 
will be FWHM$(N)\simeq 2 x_0 \sqrt{2\ln(2)p_{\rm B}(N)[1-p_{\rm B}(N)]M}$. 
On the other hand, the peak separation between 
polymers of sizes $N$ and $N+\delta N$ in a polydisperse sample
will increase as $\delta X_P(N)=M x_0 (\delta N)[dp_{\rm B}/dN]$.
Thus, resolution can be improved indefinitely by passing the polymers 
through more rows of cells. Optimal resolution is achieved
when $N\sim N_0$, for which $M\sim (\delta N/N)^2$ rows are needed in
order to resolve these two peaks. For a cell size of about 5 microns, up to 
$M=10^4$ rows should be feasible, enabling single-segment resolution up to 
$N\sim 100$ and $1\%$ resolution beyond that. Note, however, that resolution
can be further enhanced by gradient methods that are frequently implemented
in gel electrophoresis\cite{Hawcroft}, in this case by a spatially varying 
electric field or cell size.

Another major concern is the throughput, which is affected by various 
factors, including cross-sectional area, concentration and velocity of 
the polymers, and ease of specimen loading and extraction. Original 
electrolithographic designs proposed and tested by Austin and coworkers
were extremely shallow\cite{Austin,patent}, with depths comparable 
to $\ell_p$, in order to maximize hooking and to
enable individual visualization of polymers. The sieve design presented here,
on the other hand, will benefit from increased depth to at least $R_g$,
and the cells should be designed as deep as possible to increase throughput.
A stacked configuration might be considered if the mechanical stability
of posts becomes a concern.
A major bottleneck is the entrance to the sieve, since all polymers
should start from the same point, rather than a band, to achieve separation. 
This will give rise to significantly increased concentrations near the 
entrance to the sieve. Although a dilute solution is assumed 
in the calculation presented
here, separation is not limited to the dilute regime and occurs in
semidilute solutions as well. Since the mobility of polymers is significantly
higher in the absence of a gel medium, much higher concentrations can be
tolerated, therefore this bottleneck may not be as problematic as it appears. 
Furthermore, the easy and quick extraction of the separated specimens 
enables continuous operation, in which polymers are constantly added 
at the entrance and extracted from exit channels placed at the bottom 
of the sieve. Although the sieves might be expensive to produce, they
are tunable and reusable, significantly lowering
their effective cost. Thus, this technique may have major advantages 
over traditional methods for separation of large quantities.

The same technique can be applied to separation through 
centrifugation\cite{Birnie} as well.
Individual sieves shaped as pie slices, with their entrances at the apex,
can be arranged as stacked pies. The polydisperse solution can then be fed
through a tube along the rotation axis and separated objects can be collected
in containers around the circumference that rotate with the sieves. 
Virtually any separation characteristic can be achieved by modulating 
the cell size as a function of distance to the rotation axis. 

The particular design presented here is only one possibility and is
by no means an optimal geometry, although its performance is expected
to be quite satisfactory. On the other hand,
the scalable structure of the sieve makes it possible to
fully characterize device performance from a detailed modeling of
dynamics within a single cell, significantly simplifying a numerical
design effort. Furthermore, technological requirements for an experimental 
realization is well within today's capabilities, as has been demonstrated 
by Volkmuth {\em et al.}\cite{Austin}. Therefore, experimental verification
of the soundness of the general approach and a feasibility study of a
prototype device is within reach in the very near future.

This research was supported by the National Science Foundation, by the
MRSEC program through Grant No. DMR-9400396, and through Grants No.
DMR-9106237, No. DMR-9417047, and No. DMR-9416910.

\end{document}